\documentclass[12pt]{article}
\usepackage{epsfig}
\renewcommand{\u}[1]{\bar{#1}}
\newcommand{\eqn}[1]{(\ref{#1})}
\newcommand{\ft}[2]{{\textstyle{\frac{#1}{#2}}}}
\newcommand{\tr}{{\rm tr}}
\renewcommand{\H}{{\cal H}}
\begin{document}
\renewcommand{\theequation}{\thesection.\arabic{equation}}
\csname @addtoreset\endcsname{equation}{section}
\font\mybb=msbm10 at 12pt
\newcommand{\be}{\begin{equation}}
\newcommand{\ee}{\end{equation}}
\newcommand{\bea}{\begin{eqnarray}}
\newcommand{\eea}{\end{eqnarray}}
\newcommand{\dual}{\raise3pt\hbox{${\textstyle \star}$} }
\begin{titlepage}
\begin{flushright}
SU-ITP-98/02\\ KUL-TF-98/08\\ hep-th/9801206\\
\end{flushright}
\vspace{.5cm}
\begin{center}
\baselineskip=16pt
{\Large\bf CONFORMAL THEORY OF M2, D3, M5 \\
\
AND `D1+D5' BRANES }
\vskip 0.3cm
{\large {\sl }}
\vskip 10.mm
{\bf Piet Claus$^a$,  ~Renata
Kallosh$^{b}$, ~Jason Kumar$^{b}$,\\
 Paul K. Townsend$^{c,\star}$
and {}~Antoine Van Proeyen$^{a,\dagger}$ } \\
\vskip 0.8 cm
 $^a$ Instituut voor theoretische fysica, \\ Katholieke
Universiteit Leuven, B-3001 Leuven, Belgium\\[2mm]
$^b$Physics Department, \\ Stanford
University, Stanford, CA 94305-4060, USA\\[2mm]
$^c$ Institute for Theoretical Physics, University of California, \\
Santa Barbara, CA 93106, USA
\\ \vspace{6pt}
\end{center}
\vfill
\par
\begin{center}
{\bf ABSTRACT}
\end{center}
\begin{quote}
 The bosonic actions for M2, D3 and M5 branes in their own $d$-dimensional
near-horizon background are given in a manifestly $SO(p+1,2)\times SO(d-p-1)$
invariant form ($p=2,3,5$).  These symmetries result from a breakdown of $ISO(d,2)$
(with $d=10$ for D3 and $d=11$ for M2 and M5)
symmetry by the Wess-Zumino term and constraints.
The new brane actions, reduce after gauge-fixing and solving constraints
to $(p+1)$ dimensional interacting field theories with a non-linearly
realized $SO(p+1,2)$ conformal invariance. We also present an interacting
two-dimensional conformal field theory on a D-string in the near-horizon
geometry of a D1+D5 configuration.
\vfill
 \hrule width 5.cm
\vskip 2.mm
{\small
\noindent $^\dagger$ Onderzoeksdirecteur FWO, Belgium \\
\noindent $^\star$ On leave from DAMTP, Univ. of Cambridge, UK}
\end{quote}
\end{titlepage}
\section{Introduction}
It is well-known that the anti-de Sitter group is isomorphic to the conformal
group in one lower dimension. For this reason the isometry group of anti-de
Sitter space acts as the conformal group on its boundary, which is a conformally
compactified Minkowski spacetime of one lower dimension.  The degrees of freedom
at the boundary are the singleton representations of the $adS$ group (see e.g.
\cite{Fronsdal}), so singleton field theories are natural candidates for
conformal field theories. If one views the singleton fields as describing
fluctuations of the boundary of the $adS$ space then it is natural to interpret
the boundary as a physical object, i.e. a brane: this was the idea behind the
`membrane at the end of the universe' \cite{BDPS}. The connection between
singletons and branes can also be understood from the fact that the supergravity
solutions representing M2, D3 and M5 branes interpolate between the Minkowski
vacuum and a compactification to $adS$ space \cite{GT}. On the one  hand the
field modes trapped near the core are naturally identified as the worldvolume
fields of the brane. On the other hand, the core region is also the boundary
region of the $adS$ vacuum. In the past, these ideas motivated various
constructions of free (super)conformal field theories on the boundary of
$adS_{p+2}$ or, more generally, conformally flat spacetimes of the same
$S^p\times S^1$ topology \cite{inthepast,A3}. Some recent developments
\cite{m5tens,Maldacena,KKR,FF} have now brought this program back to
life.
\par
One key observation is that if a $p$-brane worldvolume is embedded appropriately
in the $adS_{p+2}$ background then it is not necessary to push the brane to the
$adS$ boundary, as was found to be necessary (for $p=2$) in \cite{BDPS}.
One way to see this is to observe that for a fixed distance from an infinite
static M2 brane (for example) the near-horizon $adS_4\times S^7$ geometry
\cite{GT} becomes a better and better approximation as the membrane number $N$
(i.e. its charge, classically) increases \cite{Maldacena}. Now, given say 1001
parallel coincident M2 branes we can move one of them away; if it is static and
remains parallel to the other 1000 membranes it will feel no force, but it will
now appear to be a `test' brane in the $adS_4\times S^7$ background of the other
1000. Since the distance of this test brane from the `source' branes is
arbitrary there is an apparent contradiction with the result of \cite{BDPS},
which we resolve in Appendix A. Here we need note only that the implication of
the above argument, i.e. that in an $adS_4\times S^7$ background there exists a
solution of the M2 brane equations that describes a static infinite flat
membrane at arbitrary $adS$ radius, is correct. The corresponding gauge-fixed
M2 brane action must be a 3-dimensional field theory with a Minkowski vacuum.
The symmetries of this theory are those of the $OSp(8|4)$ isometry supergroup of
the background, which we may now interpret as a superconformal group acting on
worldvolume fields. We thus find a 3-dimensional superconformal field theory on
the worldvolume of the M2 brane, which turns out to be an {\sl interacting}
field theory. In the limit that the M2 brane is pushed to the $adS$ boundary
we recover the free superconformal field theory on the `membrane at the end of
the universe'. The observation that the membrane need not be placed at the
boundary is therefore crucial in allowing the introduction of interactions into
(super)singleton field theories.
\par
The ideas just sketched have been used recently to make some progress towards
the construction of interacting superconformal field theories on brane
worldvolumes \cite{Maldacena,KKR,FF}, in particular on the worldvolumes of
the M2, D3 and M5 branes. The essential features shared by these cases is that
the corresponding static supergravity p-brane solution has a non-singular
degenerate Killing horizon with respect to a Killing vector field that is
timelike near spatial infinity \cite{DGT}, with the near horizon geometry being
$adS_{p+2}\times S^{d-p-2}$ \cite{GT}. In addition there is a non-vanishing
$(p+1)$-form gauge potential. The (super)conformal invariance of the action for
a test brane in this background was conjectured in \cite{Maldacena,KKR}, where
it was pointed out that the $adS$ radius of curvature $R$ is also the coupling
constant of the worldvolume field theory. More precisely, the coupling constant
is the dimensionless ratio $R/\langle r\rangle$, where $\langle r\rangle$
is the distance of the brane from the horizon, so that the weak coupling limit
$R\rightarrow 0$ (at fixed $\langle r\rangle$) is equivalent to the `end of the
universe' limit $\langle r\rangle \rightarrow\infty$ (at fixed $R$). The
conformal invariance of the action after a truncation to radial $adS$
excitations was also verified in \cite{Maldacena,KKR}. Here we shall fully
establish the conformal invariance of the bosonic actions by systematically
exploiting the fact that this symmetry follows directly from the isometries of
the background. We thereby find new (classically) conformal invariant field
theories on the worldvolumes of the M2, D3 and M5 branes.
Our method is also novel. It involves a reformulation of the (worldvolume
diffeomorphism invariant) brane action in which the $SO(d,2)$ conformal
invariance is manifest, being realized linearly as transformations in a space
of $d+2$ dimensions. There is therefore a possible connection to
a $(10+2)$ dimensional F-theory \cite{vafa} underlying the IIB branes and an
$11+2$ dimensional S-theory \cite{bars} underlying the M-branes
(see \cite{sezgin} for a recent review of supersymmetry in dimensions beyond
eleven). In our case, however, the additional dimensions serve to linearly
realize conformal invariance rather than a higher-dimensional Lorentz
invariance\footnote{It follows that our actions should not be viewed as (bosonic)
realizations of the super p-branes in spacetimes of non-Lorentzian signature
proposed in \cite{duff}.}. The {\sl manifest} conformal invariance is broken
when one attempts, by gauge-fixing and solving constraints, to express these
actions in terms of only the `physical' fields,  but the conformal invariance is
still there; it is just non-linearly realized. The choice of background is
clearly crucial to this result because symmetries of brane actions correspond to
isometries of the background. For a  flat background, for example, it has been
established \cite{m5tens} that the worldvolume  field theory is (super)conformal
only in the linearized  approximation. We understand now the conformal
symmetries of this linearized approximation as the limit $R\rightarrow
0$ of the conformal action in $adS$ background.
\par
We use essentially the same methods to find an interacting conformal invariant
(classical) field theory on the worldvolume of a D-string in the near-horizon
background of a D-string parallel to a D5-brane, we call this the `D1+D5' case.
The near-horizon geometry of various intersecting brane configurations has been
discussed previously in \cite{Hyun,KK,Skenderis,cowtown}. In our case it is
$adS_3\times S^3 \times E^4$, which makes it essentially equivalent to the near-horizon
$adS_3\times S^3$ geometry  of the $d=6$ self-dual string \cite{DGT}. Thus, the
D1+D5 case is similar to the $d=6$, $p=1$, realization of the $adS_{p+2}\times
S^{d-p-2}$ class of geometries near brane horizons. In addition, as we show
here, there is a partial restoration of supersymmetry near the horizon
(corresponding to total restoration in the context of the $d=6$ self-dual
string). This complements the already known restoration of supersymmetry near
the M2, D3 and M5 brane horizons. Thus, in each of the four cases (M2,  D3, M5 and
D1+D5) the $SO(d,2)\times SO(d-p-1)$ isometry of the background is actually the
bosonic subgroup of a supergroup. The latter are
\begin{eqnarray}
M2 :&& OSp(8|4)\nonumber\\
D3 :&& SU(2,2|4)\nonumber\\
M5 :&& OSp(6,2|4)\nonumber\\
D1+D5:&& SU(1,1|2)\times SU(1,1|2)\ .\nonumber
\end{eqnarray}
These supergroups have been indicated by Nahm \cite{nahm} as super-$adS$ and superconformal
symmetries. The last one is listed among the $adS_3$ supergroups of
\cite{GNST}, where the supersingleton representations can also be found. We
therefore expect that all the results presented here have supersymmetric
extensions in which these isometry supergroups are realized as superconformal
symmetries on the brane, but we leave this to a future investigation.
\par
In section~\ref{surface} we will consider a class of world-volume actions
in $adS_{p+2}\times S^{d-p-1}$ backgrounds. In this part there are
only scalar fields. We present an $ISO(d,2)$ symmetric form of the
action with Lagrange multiplier terms which break the symmetry down
to $SO(p+1,2)\times SO(d-p-1)$. We show how this group is realized as a rigid
extended superconformal symmetry of the gauge-fixed action.
In section~\ref{ss:M2D3M5} this general result is applied, with the necessary
modifications, to the $M2$, $D3$, $M5$ and $D1+D5$ branes.
The Wess-Zumino terms are given in each case and
the expansion of the full action about the quadratic term is given for the first
few terms (this is the action in the \lq small velocities' approximation).
Section~\ref{ss:discussion} has a short discussion of the
results and possible developments.
\par
In Appendix~\ref{app:enduniverse} we explain an apparent contradiction with \cite{BDPS},
where it was concluded that supersymmetric and static membranes only exist
when the brane is located at static infinity. We show that their actions
are included in our general action, and the difference of solutions
is due to different interpretations to the word `static', related to the
existence of more than one timelike Killing vector field.
In Appendix~\ref{app:susyD1D5} we verify the enhancement of
supersymmetry near the horizon of the D1+D5
configuration. In Appendix~\ref{app:smallv} we explain how the
Born-Infeld actions are expanded in the approximation of
small velocities.
\section{Conformal invariance `on the brane'}\label{surface}
\subsection{World-volume action}
Let $\sigma^\mu,\, (\mu=0,1,\dots,p)$ be coordinates on the $(p+1)$-dimensional
world volume $W$ of a $p$-brane, with metric $g_{\mu\nu}$ induced from a metric ${\cal G}$
on a  $d$-dimensional spacetime ${\cal M}_d$. We want to consider $p$-brane
actions of the form
\begin{equation}\label{actiona}
S= -\int_W d^{p+1}\sigma \, \sqrt{-\det g_{\mu\nu}} + \int_W A\ ,
\end{equation}
where $A$ is a $(p+1)$-form induced from a $(p+1)$-form gauge potential
${\cal A}$ on ${\cal M}_d$.  The first term is invariant under the isometries of
the background spacetime metric $g$, which are generated by killing vector
fields $\xi$. The second term will also be invariant provided that, for all
$\xi$,
\begin{equation}
{\cal L}_\xi (d{\cal A}) =0\ ,
\end{equation}
where ${\cal L}_\xi$ is the Lie derivative with respect to $\xi$.
We choose the metric on ${\cal M}_d$ to be the $SO(p+1,2)\times SO(d-p-1)$
invariant metric on $adS_{p+2} \times S^{d-p-2}$. In horospherical coordinates
for $adS$, and hyperspherical coordinates for $S^{d-p-2}$, this metric is
\begin{equation}\label{invmetric}
ds^2=\phi ^2\,dx^m\,\eta_{mn}\,dx^n+(wR)^2 \left( \frac{ d\phi }{\phi }
\right)^2 +R^2 d^2\Omega \, ,
\label{induced}\end{equation}
where $d\Omega^2$ is the $SO(d-p-1)$-invariant metric on the unit
$(d-p-2)$-sphere. The constant $R$ is clearly the radius of curvature of the
$S^{d-p-2}$ factor.
It is also proportional to the $adS$ radius of curvature $wR$ and the
constant $w$ is chosen to be
\begin{equation}
w = (p+1)/ (d-p-3)\,.    \label{winpd}
\end{equation}
We note, in passing, that this metric is
conformally flat (vanishing Weyl tensor) in those cases for which $w=1$.
For future use we note that in terms of a radial coordinate $r$ defined by
\begin{equation}\label{coordchange}
\phi = \left({r\over R}\right)^{1\over w} \ ,
\end{equation}
the metric becomes
\begin{equation}\label{nearmetric}
ds^2= dX^M\,{\cal G}_{MN}\,dX^N=\left( {r\over R}\right)^{2\over w} d {x}
^m\eta_{mn} dx^n+\left ({R\over r}\right)^2 [dr^2 + r^2d\Omega^2 ]\, .
\end{equation}
In these coordinates the infinitesimal action of the $SO(p+1,2)$ isometry group
is
\begin{eqnarray}
\label{conftrans}
\delta_{adS}(\xi) x^m &=& - \hat \xi^m (x,r) = -\xi^m(x) - (wR)^2 \left(\frac
Rr\right)^{\ft2w} \Lambda_K^m\,,\nonumber \\
\delta_{adS}(\xi) r &=& w \Lambda_D(x)\,r\,,
\end{eqnarray}
where
\begin{eqnarray}
\xi^m (x) &=& a^m + \lambda_M^{mn} x_n + \lambda_D x^m + (x^2\Lambda_K^m - 2
x^m x\cdot \Lambda_K)\,, \nonumber\\
\Lambda_D (x) &=& \ft 1d \partial_m \xi^m = \lambda_D - 2 x\cdot
\Lambda_K\ ,
\end{eqnarray}
and $a^m, \lambda_M^{mn}, \lambda_D, \Lambda_K^m$ are the constant parameters
associated with translations $P_m$, Lorentz transformations $M_{mn}$,
dilatations $D$ and special conformal transformations $K_m$.
The $adS_{p+2}$ isometries are given in terms of the parameters of the
conformal transformations in $p+1$ dimensions. More about the expressions
for $\xi^m$ and $\Lambda_D$ can be found in section 2 of \cite{m5tens}.
These transformations may be found directly by solving the Killing condition to
find the Killing vector fields $\xi$, and then computing their action on
the coordinates.
\par
Alternatively, we may proceed as follows. We introduce
a  space ${\cal S}_{d+2}$ of dimension $(d+2)$ with coordinates
\begin{eqnarray}
X^{\hat M} &=&\left \{X^{\hat m} =  \{ X^m, X^-,X^+\} , X^{\hat m'}
\right \}\nonumber\\ &&\hfill (m= 0,1, \dots, p;\ \ \hat m'=p+1, \dots, d-1)
\end{eqnarray}
and a flat metric with (mostly plus) signature $(d,2)$, i.e.
\begin{equation}
ds^2= dX^m\eta_{mn}dX^n-dX^+dX^- +  d X^{\hat m'} \delta_{\hat m'\hat n'}
dX^{\hat n'} \ ,
\end{equation}
where $\eta$ is the $(p+1)$ Minkowski metric and $\delta$ is the
$(d-p-1)$ Euclidean metric. The $adS_{p+2} \times S^{d-p-2}$ submanifold is
the intersection of the two hypersurfaces defined by the constraints
\begin{eqnarray}
T_1&\equiv & X^m \eta_{mn} X^n-X^+X^- +(wR) ^2 =0 \ , \label{defT1}\\
\nonumber\\
T_2 &\equiv & \delta_{\hat m'\hat n'} X^{\hat m'} X^{\hat n'} -  R^2  =0 \ .
\label{hyper}\end{eqnarray}
The $ISO(d,2)$ invariance of the metric on ${\cal S}_{d+2}$ is broken by these
constraints to $SO(p+1,2)\times SO(d-p-1)$. The first constraint may be solved
by setting
\begin{eqnarray}
X^-&=&\phi\,, \nonumber\\ X^m &=& \phi x^m\,,
\nonumber\\ X^+&=&\frac{(wR)^2 +\phi ^2x^2}{\phi } \,.
\label{horosphercoord}
\end{eqnarray}
The second constraint defines a $(d-p-2)$-dimensional hypersphere of radius
$R$ in ${\hbox{\mybb E}}^{d-p-1}$ and can be solved in terms of hyperspherical
coordinates\footnote{Note that the $\theta$'s do not cover the whole sphere
and we need at least two coordinate patches.  However, we will
only work in one patch.} $\theta_i$
\begin{equation}
X^{\hat m'} = \left\{ R \cos \theta_1\,,R\cos \theta_2\sin \theta_1\,,
\dots\,, R \prod_{i=1}^{d-p-2} \sin \theta_i \right\}\,.
\end{equation}
The induced $d$-metric on the submanifold defined by these
constraints is precisely the metric (\ref{invmetric}). The $SO(p+1,2)$
invariance is linearly realized on the $X^{\hat m}$ coordinates,
\begin{equation}
\delta X^{\hat m} = -\Lambda^{\hat m}_{\ \hat n} X^{\hat n}\,;\qquad
\Lambda_{\hat m\hat n} = -\Lambda_{\hat n\hat m}\,.
\end{equation}
The non-linear realization on the $(x_m,r)$ coordinates (\ref{conftrans})
may therefore be found from the relation (\ref{horosphercoord}),
by identifying
\begin{eqnarray}
a^m &= &\Lambda^m_{\ -}\,;\qquad \lambda_M^{mn} =
\Lambda^{mn}\nonumber\\
\lambda_D &=& \Lambda^+_{\ +}=-\Lambda^-{}_-\,;\qquad \Lambda_K^m = \Lambda^m_{\ +}
=\ft12\Lambda^{-m}\,.
\end{eqnarray}
\par
The above construction makes it simple to find an $SO(p+1,2)\times SO(d-p-1)$
invariant $(p+2)$-form field strength $d{\cal A}$ on ${\cal M}_d$. We start from
the following $(p+2)$-form on ${\cal S}_{d+2}$: (with $\varepsilon_{01\ldots
p+-}=\frac{1}{2}$, an anti-density as required to be constant)
\begin{equation}
\Omega_{p+2} = \frac{p+1}{(p+2)! (wR)^2}\varepsilon_{\hat
m_0\ldots \hat m_{p+2}} X^{\hat m_0}\,dX^{\hat m_1}
dX^{\hat m_2}\ldots dX^{\hat m_{p+2}} \,.  \label{OmegaWZ}
\end{equation}
This is manifestly $SO(p+1,2)$ invariant and trivially $SO(d-p-1)$
invariant. When restricted to ${\cal M}_d$ through \eqn{defT1}, it becomes
proportional to the volume
form of $adS_{p+2}$, which is closed on the constraint surface.
In fact
$\Omega_{p+2}= d{\cal A}$ on this hypersurface, where
\begin{eqnarray}
{\cal A} &=& dx^0\dots dx^p \phi^{p+1} \nonumber\\
&=& dx^0 \dots dx^p  \left ({r\over R}\right )^{p+1\over w}\, .
\label{calAp}
\end{eqnarray}
For later convenience, we combine on ${\cal M}_d$  the
horospherical coordinate $r$ and the
hyperspherical coordinates $\theta_i$ into $d-p-1$ cartesian coordinates
$X^{m'}$. The following relations hold (with $i=1,\ldots ,d-p-2$):
\begin{eqnarray}
r^2&=& X^{m'} \delta_{m'n'} X^{n'}\,,\nonumber\\
\theta_i &=& \arctan \left[ \frac {\left(\sum_{k'=i+p+1}^{d-1}
(X^{k'})^2\right)^{1/2}}{X^{(p+i)'}}\right]\,.\label{rthetatoX}
\end{eqnarray}
With the above choice of background and in these coordinates the
action (\ref{actiona}) becomes
\begin{equation}\label{actionb}
S=- \int_W d^{p+1}\sigma \left[  \sqrt{-\det g_{\mu\nu}} +  \left(\frac
rR\right)^{\ft{p+1}{w}}(\varepsilon^{\mu_0\ldots \mu_p}\partial_{\mu_0}x^0\ldots
\partial_{\mu_p}x^p) \right]\,,
\end{equation}
where
\begin{equation}
g_{\mu\nu} = \left( \frac rR\right)^{\ft 2w} \partial_\mu x^m\partial_\nu
x^n \eta_{mn} + \left( \frac Rr\right)^2 \partial_\mu X^{m'} \partial_\nu
X^{n'}\delta_{m'n'}\,.
\end{equation}
The $SO(p+1,2)\times SO(d-p-1)$ action follows, as we have essentially seen
above, from its equivalence to the {\it manifestly}  $SO(p+1,2)\times SO(d-p-1)$
action
\begin{equation}
\hat I = \hat I_D + \hat I_{WZ}+ \hat I_{LM} \,,
\end{equation}
where $\hat I_D$ is the standard Dirac $p$-brane action with a metric induced
from the flat metric on ${\cal S}_{d+2}$:
\begin{equation}
\hat I_{\rm D}  = -\int_W d^{p+1}  \sigma  \sqrt{- \det { \hat g }_{\mu\nu}}
\ ;\qquad \hat g_{\mu\nu} = \partial_\mu X^{\hat M} \partial_\nu X^{\hat N}
\eta _{\hat M \hat N}\ .   \label{Daction}
\end{equation}
The Wess-Zumino term is
\begin{equation}
\hat I_{WZ} = \int_B \Omega^B_{p+2}\ ,
\end{equation}
where $\Omega^B_{p+2}$ is the pullback of $\Omega_{p+2}$ in \eqn{OmegaWZ} to a
$(p+2)$-dimensional manifold $B$ with boundary $\partial B=W$.
The `Lagrange multiplier' action $\hat I_{LM}$
imposes the constraints $T_1=T_2=0$ via Lagrange multipliers,
\begin{equation}
\hat I_{LM} =\int_B d^{p+2}\hat \sigma \left(  \lambda^1 T_1+ \lambda^2 T_2\right)\ .
\end{equation}
Upon use of the constraints $T_1=T_2=0$ we are restricted in target
space to ${\cal M}_d$, and we may use $\Omega_{p+2}=d{\cal A}$. The action $\hat I$
thus reduces to \eqn{actionb}.
\par
In the following section we shall see how this reformulation of the standard
$p$-brane action applies to various cases of interest in M-theory and
superstring theory. First we consider its symmetries and gauge fixing
to a conformal action.
\subsection{Symmetries and gauge fixing}
The action (\ref{actionb}) is invariant under local world-volume diffeomorphisms
$\delta_{ld}(\eta)$
and under the rigid symmetries determined by the Killing equation,
i.e. the $adS$ symmetries $\delta_{adS}(\xi)$ and sphere isometries.
We shall fix the local gauge invariance by the `physical', or `static', gauge
condition
\begin{equation}\label{choice}
x^\mu (\sigma)\equiv \delta^\mu{}_m \; x^m (\sigma) = \sigma^\mu\,.
\end{equation}
The gauge-fixed action is then
\begin{equation}\label{fix}
S_{g.f} = S\big|_{x=\sigma}\, .
\end{equation}
The gauge-fixing breaks not only the local diffeomorphisms (as it is designed to
do) but also the rigid $adS$ symmetry. However, the action will still be
invariant under some combination of $adS$ symmetry plus `compensating'
local diffeomorphisms $\delta_{ld}(\eta)$, for a particular `rigid' parameter
$\eta(\sigma)$ defined by the requirement
\begin{equation}
[\delta_{adS}(\xi) + \delta_{ld}(\eta (\sigma))] x^\mu\big|_{x=\sigma} =0\ ,
\end{equation}
which yields
\begin{equation}
\eta^\mu (\sigma)  = \hat \xi^\mu (\sigma , r(\sigma))\ .\label{decomp}
\end{equation}
We thus find that the gauge-fixed action (\ref{fix}) is
invariant under the following conformal transformations
\begin{eqnarray} \label{confX1}
\delta_C X^{m'} &=& \hat \xi^\mu(\sigma,r(\sigma)) \partial_\mu X^{m'} + w
\Lambda_D(\sigma) X^{m'}\\
&=&\xi^\mu (\sigma) \partial_\mu X^{m'} + w \Lambda_D(\sigma) X^{m'} +
(wR)^2 \Lambda^\mu_K \left(\frac Rr\right)^{\ft2w} \partial_\mu X^{m'}\,,
\nonumber\end{eqnarray}
which are of the general form (2.4) in
\cite{m5tens}, identifying $w$ as the Weyl weight of $X^{m'}$, and
the special conformal transformations acquire an extra part
\begin{equation}
 k_\mu X^{m'}=(wR)^2 \left({R\over r }\right)^{2/w} \partial_\mu X^{m'} \ .
\end{equation}
For completeness we give the $SO(d-p-1)$ \lq R-symmetry' acting linearly on
the $X^{m'}$ as
\begin{equation}
\delta_R X^{m'}= -\Lambda^{m'}_{\ n'} X^{n'}\,,\qquad \Lambda_{m'n'} = -
\Lambda_{n'm'}\,.\label{SOX1}
\end{equation}
In arriving at this result we have implicitly assumed that the gauge condition
$x^\mu=\sigma^\mu$ is valid {\it globally}, rather than just locally. This
assumption restricts the possible worldvolume field configurations to some set
that includes an infinite static brane aligned with the cartesian $x^\mu$ axes.
The worldvolume of this infinite static brane will be identified as the
Minkowski vacuum of the gauge-fixed action. For this worldvolume vacuum to be
stable it must be a solution of the worldvolume field equations (the `branewave'
equations) but it is only in very special backgrounds that such an infinite
static brane solution exists. In the backgrounds considered here it was once
thought that an infinite static brane could solve the branewave equations only
if placed at the $adS$ boundary at $r=\infty$ \cite{BDPS}. This is indeed a
solution but not the only one: in the coordinates used here the
`vacuum' configuration
\begin{equation}
x^\mu =\sigma^\mu, \qquad r= \langle r\rangle,
\end{equation}
with constant position on the $(d-p-2)$-sphere, is a solution for {\it any}
value of $\langle r\rangle$.
\section{M2, D3, M5 and D1+D5 conformal field theories}\label{ss:M2D3M5}
We now have most of the ingredients needed to demonstrate the conformal
invariance of the bosonic M2, D3, M5 and D1+D5 brane actions in their own
near-horizon backgrounds.
The space-time metric and the $(p+1)$-form of the brane
configuration with 1/2 of unbroken supersymmetry are
described in terms of a harmonic function
\begin{equation}
 H_{\rm brane} = 1+ \left ({R\over r}\right )^{p+1\over w}\ ,
\hskip 2 cm
r^2=X^{m'}\delta_{m'n'} X^{n'}  \ ,
\label{harm}\end{equation}
which solves the Laplace equation in the space of
$d-p-1$ transverse directions $X^{m'}$ to the brane. The
metric is ($m= 0,1,
\dots,p; \;
m' =p+1, \dots ,d-1$):
\begin{equation}
ds^2_{\rm brane}=H_{\rm brane} ^{-{2\over p+1}} dx^m\eta_{mn}dx^n + H_{\rm
brane}
^{2w\over p+1}
dX^{m'}\delta_{m'n'}dX^{n'} \ .
\label{brane}\end{equation}
The brane space-time is asymptotically flat at $r
\rightarrow
\infty $. The
horizon of the brane in these isotropic coordinates is
at $r
\rightarrow 0$.
Very close to the horizon at $r \rightarrow 0$ the first
term in the harmonic function (\ref{harm}) can be
neglected comparative to the second term. This gives a
near-horizon geometry as in \eqn{nearmetric} and a $(p+1)$-form
as in \eqn{calAp} \cite{GT}.
It has been pointed out in \cite{Maldacena} that one can
interpret the parameter $R$ for each of the
configurations above as follows. For the M2 brane $R^6=
2^5 \pi^2 N l_{P}^6$, for the D3 brane $R^4= 4\pi
gN\alpha'^{2}$, and for M5 $R^3=
\pi N l_P^3$ and $N$ can be interpreted as the number of parallel
branes. It is possible therefore to think of the
$adS_{p+2}\times S^{d-p-2}$ geometry as coming either as
a result of considering a small $r$ (near horizon)
approximation or a large $N$ (many branes with unit
charges) approximation. Note also that since the
curvature near the horizon depends on N as a negative
power, large $N$ is required for supergravity to be valid.
An additional interpretation was suggested in
\cite{Hyun,Skenderis,CLLPS}, where a special duality transformation
was introduced to remove the constant from the harmonic
function. It is this metric \eqn{nearmetric},
which gives the target space metric ${\cal G}$ that is pulled
back to the brane worldvolume to lead to the metric $g_{\mu\nu}$ used
in section~\ref{surface}.
\par
The above formalism with the relation \eqn{winpd} can be applied in
all the cases in table~\ref{tbl:branesdpw}.
\begin{table}[h]\begin{center}\begin{tabular}{||l|c|c|c||}\hline
                              & d & p & w    \\ \hline
M2                            & 11& 2 & 1/2  \\
M5                            & 11& 5 & 2    \\
D3                            & 10& 3 & 1    \\
Self-dual string              & 6 & 1 & 1    \\
Magnetic string               & 5 & 1 & 2    \\
Tangerlini black hole         & 5 & 0 & 1/2  \\
Reissner-Nordstr\"om black hole & 4 & 0 & 1   \\
\hline
\end{tabular}\end{center}\caption{Solutions with an $adS_{p+2}\times S^{d-p-2}$ geometry.}
\label{tbl:branesdpw}\end{table}
The first three cases have in common the feature that $w$
is the canonical Weyl weight of a world-volume scalar field, i.e.
\begin{equation}
w = {1\over2}(p-1) \qquad {\rm (M2,\ D3,\ M5)}\,.
\end{equation}
The other ones can be obtained as intersecting branes and have as
such been mentioned in \cite{KK,Skenderis,cowtown}.
Then the resulting geometry has as extra factors Euclidean manifolds.
Apart from M2, D3 and M5, we shall treat here the $d=6$ case as the intersection
`D1+D5'. The black hole cases have some interesting special features which will be
discussed in a separate article \cite{us}.
\par
The world-volume actions are in all those cases similar to those
treated in section~\ref{surface}, but we will need modifications of
the Wess-Zumino (WZ) terms, and additions due to the presence of a
world-volume vector for D3 and a world-volume antisymmetric tensor
for M5.
\subsection{M2 brane}
The near-horizon background of the M2 brane is the $adS_4\times S^7$
solution as mentioned before, i.e. \eqn{nearmetric} and \eqn{calAp}
with $d=11$, $p=2$ and $w=\ft12$. The results found there apply without change.
The gauge-fixed M2 brane action can be expanded in a power series in $R$,
which is appropriate in the small velocities approximation. The $R$-independent
term is just a free conformal field theory so $R$ is the coupling constant, as
pointed out in \cite{Maldacena, KKR}. Details of this expansion can be found
in Appendix~\ref{app:smallv}. The action up to fourth order in derivatives of fields is
\begin{eqnarray}
S_{\rm M2} = - \int d^3\sigma \hspace{-6mm}&& \Big\{ \frac12
\partial^{\u\mu} X^{m'} \partial_{\u\mu} X_{m'}\nonumber\\ && -
\frac14 \left( \frac Rr \right)^6[\partial^{\u\mu} X^{m'}
\partial_{\u\nu} X_{m'} \partial^{\u \nu} X^{n'} \partial_{\u\mu} X_{n'}
- \frac12
(\partial^{\u\mu} X^{m'} \partial_{\u\mu} X_{m'})^2 ]
\nonumber\\ &&+ \dots \Big\}\,. \label{smallvM2}
\end{eqnarray}
The indices ${\u\mu}$ are barred to indicate
contraction with the {\sl flat} metric $\eta_{\u\mu\u\nu}$.  Note
that the truncation to $R=0$ preserves the conformal symmetry and
gives a free scalar multiplet in 3 dimensions.
\subsection{D3 brane}
The near-horizon geometry of the D3 brane is the $adS_5\times S^5$ IIB
supergravity solution $(w=1)$
\begin{eqnarray}
ds^2 &=& \left( \frac rR \right)^2 dx^mdx^n\eta_{mn} + \left(\frac
Rr\right)^2[dr^2 + r^2 d\Omega^2]\,, \nonumber \\
{\cal A}_4 &=& \left(\frac rR\right)^4 dx^0dx^1dx^2dx^3 \nonumber\\
&& + 4 R^4 \sin^4\theta_1
\sin^3\theta_2 \sin^2\theta_3 \sin\theta_4 \theta_5 d\theta_1 d\theta_2 d\theta_3
d\theta_4\,.
\end{eqnarray}
Note that the latter is not unique, e.g. the last term could also be
chosen as $$ 4 R^4 \sin^4\theta_1
\sin^3\theta_2 \sin^2\theta_3 \cos\theta_4 d\theta_1 d\theta_2 d\theta_3
d\theta_5\ .$$
This is similar to, but not quite the same as, the $d=10, p=3$ case of the
previous section. The difference arises because the 5-form field strength of
${\cal A}_4$ is now required to be self-dual. The WZ term must therefore be
changed accordingly.  We can accomplish this in a manifestly $SO(4,2)\times
SO(5)$ invariant way as follows.  We introduce the additional manifestly
$SO(6)$-invariant 5-form
\begin{equation}
\tilde\Omega_5 = {4\over 5!R^2} \varepsilon_{\hat m_0' \hat m_1'\dots
\hat m_5'}X^{\hat m_0'}dX^{\hat m_1'}\dots dX^{\hat m_5'}\ .
\end{equation}
On the hypersurface $T_2=0$ this is proportional to the volume form on
$S^5$, so it is
the hodge dual\footnote{The proportionality constant between $\Omega_5$ and
the volumeform on $adS_5$ is the same as between $\tilde \Omega_5$ and the
volume form on $S^5$.} in ${\cal M}_d$ of the restriction
to the hypersurface $T_1=0$ of $\Omega_5$. It follows that the 5-form
\begin{equation}
\Omega_{\rm D3} = \Omega_5 + \tilde \Omega_5
\end{equation}
is the self-dual field strength of ${\cal A}_4$ on the constraint surface.
The D3 WZ term is therefore obtained from the one of the previous section
by the replacement $\Omega_5 \rightarrow \Omega_{D3}$.  A modification is
also needed to the `Dirac' action to incorporate the world-volume
Born-Infeld 1-form $V$.  We take
\begin{equation}
\hat I_D \rightarrow I_{DBI} \equiv - \int d^4  \sigma
 \sqrt{- \det (g + F) } \,,   \label{BIaction}
\end{equation}
where $g$ is the metric induced from the flat metric on ${\cal S}_{d+2}$ and
$F= dV$, which is inert under
isometries of $g$ (note that the NS 2-form vanishes in the background that
we consider).
The gauge-fixed action is then invariant under (\ref{confX1}), (\ref{SOX1})
and
\begin{eqnarray}
\delta_C F_{\mu\nu} &=& \hat \xi^\rho(\sigma,r) \partial_{\rho} F_{\mu\nu} - 2
\partial_{[\mu} \xi^\rho(\sigma,r) F_{\nu]\rho}\nonumber\\
&=& \xi^\rho(\sigma) \partial_\rho F_{\mu\nu} - 2 \Lambda_M(\sigma)^{\rho}_{\ \,
[\mu} F_{\nu]\rho} + 2 \Lambda_D (\sigma) F_{\mu\nu} \nonumber\\
& & + R^2\Lambda_K^\rho \left[ \left(\frac Rr\right)^2 \partial_{\rho} F_{\mu\nu} - 2
\partial_{[\mu} \left(\frac Rr\right)^2 F_{\nu]\rho}\right]\,.
\label{dconfF}
\end{eqnarray}
This is again of the general form of conformal transformations with
e.g. on the vector $V_\mu$ the extra transformation
\begin{equation}
k_\nu V_\mu=R^2 \left( \frac Rr\right)^2
\partial_\nu V_\mu +  R^2 \partial_{\mu} \left( \frac Rr
\right)^2 V_\nu\ .
\end{equation}
This drops out for $R=0$ and in that limit we get the free action of
the bosonic part of a $N=4$ vector multiplet in 4 dimensions.
\par
The D3 brane gives a 4-dimensional interacting conformal
theory of a vector supermultiplet which can again be expanded as a power
series in $R$ or, equivalently, in powers of derivatives of fields.  The
expansion to fourth order in derivatives is
\begin{eqnarray}
S_{\rm D3} = - \int d^4\sigma\hspace{-6mm}&& \Big\{\frac12
\partial^{\u
\mu} X^{m'}
\partial_{\u \mu}
X_{m'} + \frac14 F^{\u\mu\u\nu} F_{\u\mu\u\nu}
\nonumber\\ && -
\frac14 \left( \frac Rr \right)^4[\partial^{\u \mu} X^{m'}
\partial_{\u\nu} X_{m'} \partial^{\u\nu} X^{n'} \partial_{\u\mu} X_{n'}
- \frac12 (\partial^{\u\mu} X^{m'} \partial_{\u\mu} X_{m'})^2 ]
\nonumber\\
&& - \frac18 \left( \frac Rr\right)^4
[F^{\u\mu\u\nu}F_{\u\nu\u\rho}F^{\u\rho\u\sigma}F_{\u\sigma\u\mu}
-
\frac14(F^{\u\mu\u\nu} F_{\u\mu\u\nu})^2] \nonumber\\
&& + \frac12 \left( \frac Rr \right)^4 [\partial^{\u\mu}
X^{m'}
\partial_{\u\nu} X_{m'} F^{\u\nu\u\rho} F_{\u\rho\u\mu}
+ \frac14 \partial^{\u\mu} X^{m'} \partial_{\u\mu}
X_{m'} F^{\u\nu\u\rho} F_{\u\nu\u\rho} ]\nonumber\\
&&+4 R^4 \sin^4 \theta_1 \sin^3 \theta_2 \sin^2 \theta_3 \sin \theta_4 \theta_5
\varepsilon^{\u\mu\u\nu\u\rho\u\sigma} \partial_{\u\mu}\theta_1
\partial_{\u\nu} \theta_2\partial_{\u\rho} \theta_3 \partial_{\u\sigma}
\theta_4\nonumber\\
&&+\dots \Big\} \,,  \label{smallvD3}
\end{eqnarray}
where $(X^{m'}, r, \theta_i)$ are related through \eqn{rthetatoX}.
\subsection{M5 brane}
The near-horizon geometry of the M5 brane\footnote{For M5 we will use the
formulation of \cite{5b}, following the notation and conventions of \cite{m5tens}.} is the
$adS_7\times S^4$ solution of D=11 supergravity $(w=2)$
\begin{eqnarray}
ds^2 &=&\left(\frac rR\right)dx^mdx^n\eta_{mn} + \left(\frac
Rr\right)^2[dr^2 + d\Omega^2] \nonumber \\
{\cal A}_3 &=& -3 R^3 \sin^3 \theta_1  \sin^2 \theta_2 \sin \theta_3
\theta_4
d\theta_1 d\theta_2 d\theta_3  \ .  \label{M5soln}
\end{eqnarray}
In order to cast this into the $d=11,p=5$ case of the previous section we need
to give the solution for ${\cal A}_3$ in terms of its magnetic dual
potential ${\cal A}_6$.
In general, the two are related by $d{\cal A}_6 = \dual
d{\cal A}_3 + {\cal A}_3\wedge d{\cal A}_3$,
but the second term on the right hand side vanishes for the particular
${\cal A}_3$ of the above solution.  The dual 6-form potential therefore
satisfies
$d{\cal A}_6 = \dual d{\cal A}_3$.  We can choose it be
\begin{equation}
{\cal A}_6 = \left(\frac rR\right)^3 dx^0 dx^1 \dots dx^5\ .
\end{equation}
We see from this that the $adS_7\times S^4$ background is of the form
assumed in section~\ref{surface}.  The WZ term includes an integral over the
6-form potential, which can be dealt with as explained previously, via the
introduction of the 7-form $\Omega_7$.
\par
However, there is another contribution to the WZ term involving the
world-volume 3-form field strength $\H$, which we will introduce on
$B$, subject to the Bianchi identity
\begin{equation}
d\H + d{\cal A}^B_3 =0\, .
\end{equation}
We shall consider $\H$ to be an {\sl independent} 3-form field
and we shall introduce its Bianchi `identity'
by means of a 3-form Lagrange multiplier $K_3$.  Specifically, we add to
the action the term
\begin{equation}
I_{LM}' = \int_{B} K_3 \wedge (d\H + \tilde\Omega_4^B)
\end{equation}
where $\tilde\Omega_4^B$ is the pullback to the space $B$ of
\begin{equation}
\tilde \Omega_4 = {3\over 4!R^2} \varepsilon_{\hat n'\hat
m_1'\dots\hat m_4'} X^{\hat n'} dX^{\hat m_1'}\dots dX^{\hat m_4'}\,.
\end{equation}
\par
The Lagrange multiplier $K_3$ imposes the constraint $d\H = -
\tilde\Omega_4^B$, which requires $\tilde\Omega_4^B$ to be closed.  It is
sufficient
that this is true on the constraint surface $T_1=T_2=0$, which it is
because on this surface $\tilde\Omega_4$ is the hodge dual of $\Omega_7$ (see
\eqn{OmegaWZ}) and
hence equal to $d{\cal A}_3$.  The M5 brane WZ term can now be written in a
manifestly $SO(6,2)\times SO(5)$ invariant way as the integral over $B$
of the 7-form
\begin{equation} \Omega_{M5} = \Omega_7^B - {1\over2}
\H\wedge \tilde \Omega_4^B\ .
\end{equation}
On the constraint surface (and imposing the constraint
$d\H=\tilde\Omega_4^B$) we have
\begin{eqnarray}
&&\Omega_7=d{\cal A}_6\ ;\qquad \tilde\Omega_4^B=d{\cal A}_3=-d{\cal
H} \ ;\qquad
{\cal A}_3\wedge d{\cal A}_3=0  \nonumber\\
&&\Rightarrow\qquad \Omega_{M5}=
d\left( {\cal A}_6^B+\ft12{\cal H}\wedge {\cal A}_3^B\right) \ ,
\end{eqnarray}
so that the WZ term
reduces, locally, to an integral over the 6-dimensional boundary of $B$.
It is, in fact, just the WZ term of the M5 brane action of \cite{5b}.
The $SO(6,2)\times SO(5)$ invariance is now manifest if we take $\H$, and
the Lagrange multiplier $K_3$ to be invariant (as we may do since their
transformations must be specified independently of those of all other
fields).
A modification to the `Dirac' action relative to the
previous section is also required.
To obtain it we proceed as follows: we define \cite{5b}
\begin{eqnarray}
u_\mu &=& \partial_\mu a\,, \qquad v_\mu =
\frac{u_\mu}{\sqrt{u^2}}\,, \qquad u^2 = u_\mu \hat g^{\mu\nu} u_\nu\,,\nonumber\\
\H^*_{\mu\nu\rho} &=& {\sqrt{-\det\hat  g}\over 6}
\varepsilon_{\mu\nu\rho\sigma\lambda\eta} \hat g^{\sigma\tau} \hat g^{\lambda\kappa}
\hat g^{\eta\phi}
\H_{\tau\kappa\phi}\,,\nonumber\\
\H_{\mu\nu} &=& v^\rho
\H_{\mu\nu\rho}\,,\qquad \H^*_{\mu\nu} = v^\rho \H^*_{\mu\nu\rho}\,,
\end{eqnarray}
where $a$ is the \lq PST' auxiliary field, and $\hat g$ is the
induced world-volume metric from the flat metric in ${\cal S}_{d+2}$.
\par
We must then make the replacement
\begin{equation}
\hat I_D \rightarrow \hat I_{BLNPST} \equiv -\int d^6 \sigma \left[\sqrt{- \det (\hat g +
i \H^*) } + \ft14 \sqrt{-\det \hat g}\ \H^{*\mu\nu} \H_{\mu\nu}\right]
\,.  \label{M5action}
\end{equation}
This is again manifestly
$SO(6,2)\times SO(5)$ invariant.  The Bianchi identity can be
solved for a two form potential $B_{\mu\nu}$, i.e.
\begin{equation} \H = dB - {\cal A}^B_3\,.
\end{equation}
Note that $\tilde \Omega_4$ is invariant under $SO(5)$, but an
explicit solution for ${\cal A}_3$ of $\tilde \Omega_4=d{\cal A}_3$
is not invariant. E.g. the one we gave in \eqn{M5soln} is not
invariant under a shift of $\theta_4$. But we should have that
\begin{equation}
\delta_R(\Lambda) {\cal A}_3 = dC_2(\Lambda)\,,
\end{equation}
for some 2-form $C_2$. The requirement that $\H$ is invariant under the background
isometries forces the potential $B$ to transform under $SO(5)$ as
\begin{equation}
\delta_R(\Lambda) B = C_2(\Lambda)\ ,
\end{equation}
(up to a gauge transformation). Therefore the algebra of isometries
could acquire central charges. Indeed, from descent equations
\begin{equation}
\delta_R(\Lambda') C_2(\Lambda) -
\delta_R(\Lambda) C_2(\Lambda') = C_2\left([\Lambda,\Lambda'] \right) +
dC_1(\Lambda,\Lambda')\ .
\end{equation}
It is possible that $C_1$ vanishes, and with
 the choice of ${\cal A}_3$ in (3.12) partial calculations indicate that
 a $C_2$ exists with vanishing $C_1$.
\par
It is straightforward to compute the conformal transformations from the
decomposition rule \eqn{decomp}. The transformation of the fields
appearing in the action, can be deduced from the following.
Besides \eqn{confX1} and \eqn{SOX1}, we have
\begin{eqnarray}
\delta_C a &=& \hat \xi^\mu(\sigma,r) \partial_\mu a = \xi^\mu(\sigma) \partial_\mu a + 4R^2
\Lambda^\mu_K \left(\frac Rr\right) \partial_\mu a\,,\nonumber\\
\delta_C B_{\mu\nu} &=& \hat \xi^\rho(\sigma,r) \partial_{\rho} B_{\mu\nu} - 2
\partial_{[\mu} \xi^\rho(\sigma,r) B_{\nu]\rho}\nonumber\\
&=& \xi^\rho(\sigma) \partial_\rho B_{\mu\nu} - 2 \Lambda_M(\sigma)^{\rho}_{\ \,
[\mu} B_{\nu]\rho} + 2 \Lambda_D (\sigma) B_{\mu\nu} \nonumber\\
& & + R^2 \left[ \left(\frac Rr\right)^2 \partial_{\rho} B_{\mu\nu} - 2
\partial_{[\mu} \left(\frac Rr\right)^2 B_{\nu]\rho}\right]\,.
\end{eqnarray}
The M5 brane gives a 6-dimensional conformal theory of a tensor supermultiplet.
The bosonic action in the same approximation as before is
\begin{eqnarray}
S_{\rm M5} = - \int d^6\sigma \hspace{-6mm} &&\Big\{\frac12
\partial^{\u
\mu} X^{m'}\partial_{\u \mu}X_{m'} + \frac12 \H^{*\u\mu\u\nu}
\H_{\u\mu\u\nu}^-
\nonumber\\
&& - \frac14 \left( \frac Rr \right)^3[\partial^{\u \mu}
X^{m'}
\partial_{\u\nu} X_{m'} \partial^{\u\nu} X^{n'} \partial_{\u\mu} X_{n'}
- \frac12 (\partial^{\u\mu} X^{m'} \partial_{\u\mu} X_{m'})^2 ]
\nonumber\\
&& - \frac18 \left( \frac Rr\right)^3
[\H^{*\u\mu\u\nu}\H^*_{\u\nu\u\rho}\H^{*\u\rho\u\sigma}\H^*_{\u\sigma\u\mu}
-
\frac14(\H^{*\u\mu\u\nu} \H^*_{\u\mu\u\nu})^2] \nonumber\\
&& + \frac12 \left( \frac Rr \right)^3 [\partial^{\u\mu}
X^{m'}
\partial_{\u\nu} X_{m'} \H^{*\u\nu\u\rho} \H^*_{\u\rho\u\mu}
+ \frac14 \partial^{\u\mu} X^{m'} \partial_{\u\mu}
X_{m'}
\H^{*\u\nu\u\rho}
\H^*_{\u\nu\u\rho} ]\nonumber\\
&&+\frac12\left(\frac Rr\right)^3[v_{\u\rho}
\partial^{\u\rho} X^{m'}
\partial^{\u\sigma} X_{m'} v_{\u\sigma}
\H^{*\u\mu\u\nu}\H_{\u\mu\u\nu}^-
\nonumber\\
&&\hspace{2cm}- \frac12 v_{\u\rho} \partial^{\u\rho} X^{m'}
\partial^{\u\sigma} X_{m'}
\H_{\u\sigma\u\mu\u\nu} \H^{*\u\mu\u\nu}]\nonumber\\
&&+ \frac14 R^2 \sin^3\theta_1 \sin^2 \theta_2 \sin                                         \theta_3  \theta_4
\varepsilon^{\u\mu\u\nu\u\rho\u\sigma\u\lambda\u\tau}
\H_{\u\mu\u\nu\u\rho} \partial_{\u\sigma} \theta_1 \partial_{\u\lambda} \theta_2
\partial_{\u\tau} \theta_3\nonumber\\
&&+\dots \Big\}\,.\label{smallvM5}
\end{eqnarray}
\subsection{D1+D5}
The self-dual string near-horizon
geometry is $adS_3\times S^3$ which is very similar to the $adS_3\times S^3
\times E^4$ near-horizon geometry of the solution representing a D1-brane
embedded within a D5-brane\footnote{The
near horizon behaviour and enhancement of supersymmetry
of various intersecting and overlapping branes was
studied in \cite{KK} and \cite{Skenderis}. We derive the
enhancement of supersymmetry for D1+D5 in Appendix B.}.
In IIB supergravity in the string frame, this solution
is given by
\bea
ds^2 _{\rm brane}&=& (H_1 H_5 )^{-{1\over 2}} (-(dx^0)^2 +(dx^5) ^2 ) +
H_1
^{1\over 2} H_5
^{-{1\over 2}} ((dx^1) ^2 + ... + (dx^4) ^2) \nonumber\\
&&+ (H_1 H_5)^{1\over 2} ((dx^6) ^2 + ... + (dx^9) ^2)\nonumber\\
{\cal F}_3&=&\Omega_3+\tilde\Omega_3\nonumber\\
\Omega_3&=&dx^0\,dx^5\,dH_1^{-1}\ ;\qquad
\dual\tilde\Omega_3
=dx^0\,dx^1\,dx^2\,dx^3\,dx^4\,dx^5\,dH_5^{-1}\nonumber\\
H_1 &=& 1 + \left ({c_1 \over r} \right )^2\ ;
\qquad
H_5 = 1+\left ({c_5 \over r} \right )^2\,,
\qquad
e^{-2\phi} = H_5 H_1 ^{-1} \ .
\eea
This solution was found in \cite{CM} by performing an S-duality rotation of
 NS-NS 1+5 solution \cite{tsey}. For the case of
the  string+fivebrane solution in NS-NS
form it was shown  that the string action in the conformal gauge in
the near-horizon  region reduces to the direct product of 2-dimensional
conformal theories: $SL(2,R)$ and $SU(2)$ WZW
theories and a free 5-torus theory. The relation between this conformal theory
and the one we discuss here is still to be understood.
\par
In what follows we will use $c_1=c_5=R$, i.e. $H_1=H_5$.
Note that in the near-horizon regime, the solution
assumes the form of $adS_3 \times S^3 \times
E^4$  ($r^2= (x^6)^2+ \dots + (x^9)^2$) .
\begin{eqnarray}\label{onefive}
ds^2 &=& \left(\frac rR\right)^2 [-(dx^0)^2 +(dx^5) ^2] + \left(\frac Rr\right)^2 [dr^2
+ r^2 d\Omega^2]\nonumber\\ && + [(dx^1) ^2 + ... + (dx^4) ^2]\,, \nonumber\\
{\cal A}_2 &=& \left( \frac rR\right)^2 dx^0dx^5 + 2 R^2 \sin^2\theta_1 \sin \theta_2
\theta_3d\theta_1d\theta_2\ ;
\qquad e^{-2\phi}_{\rm hor}= 1 \,.
\end{eqnarray}
This solution is of the form $adS_3\times S^3 \times E^4$, i.e. the product of $E^4$
with the submanifold defined in section~\ref{surface}
for $d=6$ as the intersection of two
hypersurfaces $T_1=T_2=0$ in a flat space of signature (6,2). Equivalently, we
may define it as a submanifold of space of signature (10,2) subject to the same
($d=6$) constraints (which are therefore independent of the
$E^4$ coordinates).
\par
Let $g$ be the metric on the worldsheet of a D-string
induced from the flat metric on this (10,2) space.  Since the dilaton is a
constant in our chosen background, the DBI part of the action is simply
\begin{equation}
S_{DBI} = \int d^2\sigma \sqrt{-\det (g_{\mu\nu} + F_{\mu\nu})},
\label{SDBI}
\end{equation}
where $F=dV$ is the field strength of the Born-Infeld 1-form potential $V$
(which we take to be inert under isometries of the background).
This action is manifestly $SO(2,2)\times SO(4)$
invariant (and also manifestly invariant under the $ISO(4)$ isometry group of
$E^4$), which remains from  the original $ISO(6,2)$ (or $ISO(10,2)$) symmetry
 by the existence of a hypersurface.
 It was shown in \cite{BGS} that \eqn{SDBI} in 2 dimensions
has more symmetries. An infinite dimensional extension of the adS
isometries (2.7), written collectively as $\delta X^M= h^M(X)$, is
provided by
\begin{eqnarray}
\delta X^M&=& h^M(X)\, \lambda({\cal F})\nonumber\\
\delta V_\mu &=& -\lambda'({\cal F}) \sqrt{g}\,\epsilon_{\mu\nu}H^\nu
(1+{\cal F}^2)
\end{eqnarray}
where
\begin{eqnarray}
H^\mu&=&g^{\mu\nu}\left( \partial_\nu X^M\right)  g_{MN} h^N(X)
\nonumber\\
F_{\mu\nu}&=& \partial_\mu V_\nu-\partial_\nu V_\mu=\epsilon_{\mu\nu}F
\qquad\mbox{or}\qquad F=-\ft12\epsilon^{\mu\nu}F_{\mu\nu}\nonumber\\
{\cal F}&=& \frac{F}{\sqrt{g}}\ ;\qquad g=\det |g_{\mu\nu}|\ .
\end{eqnarray}
The arbitrary function $\lambda$ thus provides a sort of
Ka\v{c}-Moody extension of the isometry group. Note that this is not
the Virasoro extension. The Virasoro Killing vector of two dimensions
is not a Killing vector of the 3-dimensional adS metric
(only its $SU(1,1)$ part is). We do not yet see an application of
this extension, and will thus omit it henceforth.
\par
The WZ term of the D-string is (in the above background) the integral of
$d{\cal A}^B_2$, the pullback of ${\cal A}_2$,
over a 3-dimensional manifold $B$ having the string worldsheet as its
boundary. The WZ term is therefore $SO(2,2)\times SO(4)$ invariant if the
3-form $d{\cal A}_2$ is. Note that $d{\cal A}_2$ is actually a 3-form on $adS_3\times S^3$. In
this context it is self-dual and its $SO(2,2)\times SO(4)$ invariance may
therefore be made manifest in the same way that we made manifest the
$SU(2,2)\times SU(4)$ invariance of the D3 brane WZ term. Specifically, we
begin by defining
\begin{equation}
\Omega_3 = {1\over 3R^2} \varepsilon_{\hat m\hat n\hat p\hat q} X^{\hat m} dX^{\hat n}
dX^{\hat p} dX^{\hat q}\,,\qquad
\tilde\Omega_3  = {1\over 3R^2} \varepsilon_{\hat m'\hat n'\hat p'\hat q'} X^{\hat m'}
dX^{\hat n'} dX^{\hat p'} dX^{\hat q'}\,,
\end{equation}
where $X^{\hat m}$ are coordinates on the (2+2) dimensional space in which $adS_3$ is
defined as the hypersurface $T_1=0$, and $X^{\hat m'}$ are the coordinates on
the 4-dimensional space in which $S^3$ is defined as the hypersurface
$T_2=0$. We then take the WZ term to be the integral over $B$ of the
3-form
\begin{equation}
\Omega_{D1} = \Omega_3 + \tilde\Omega_3\,.
\end{equation}
On the constraint hypersurface this reduces to $d{\cal A}_2$ of (\ref{onefive}).
Thus, the addition to the action of a Lagrange multiplier term for the constraints
yields a manifestly $SO(2,2)\times SO(4)$ form of the D-string action in the
D1+D5 background. The gauge-fixed non-linear action,
\begin{equation}
S_{D1} = -\int d^2\sigma \left(\frac rR\right)^2\left(\sqrt{-\det( g + F )} - 1
\right) + 2 R^2 \sin^2\theta_1 \sin\theta_2\theta_3 \varepsilon^{\mu\nu}
\partial_\mu\theta_1 \partial_\nu  \theta_2\,,
\end{equation}
where
\begin{equation}
g_{\mu\nu} = \eta_{\mu\nu} + \left(\frac Rr\right)^4 \partial_\mu X^{m'}
\partial_\nu X_{m'} + \left(\frac Rr\right)^2 \partial_\mu X^{\tilde m}
\partial_\nu X_{\tilde m}
\end{equation}
and $X^{\tilde m}$ are the coordinates of $E^4$, is invariant under
the conformal transformations given in \eqn{confX1} with $w=1$ for
$X^{m'}$ and with $w=0$ for $X^{\tilde m}$, and for the vector under
\eqn{dconfF}.
\section{Discussion}  \label{ss:discussion}
M-theory and string theory $p$-brane solutions with non-singular
Killing horizons are typically asymptotic near the horizon to spaces of the form
$adS_{p+2}\times S^{d-p-2}$ . Examples are the M2 and M5 branes ($d=11$) and
the IIB D3 brane ($d=10$). In these cases we have shown that a test M2, D3 or
M5 brane in its own near-horizon geometry has a conformal symmetry inherited
from the isometries of the anti-de Sitter background. Upon fixing the
world-volume diffeomorphism invariance these theories yield interacting
conformal invariant field theories in 3,4 and 6 dimensions, respectively. There
is a single coupling constant proportional to the $adS$ radius of curvature $R$.
Since $r$ is dimensionful, worldvolume vacua with $r\ne0$ spontaneously
 break conformal invariance and are therefore degenerate. Each vacuum
solution corresponds to an infinite planar brane at a distance
$\langle r\rangle$ from the horizon, where $r$ is the Nambu-Goldstone field of
spontaneously broken conformal invariance\footnote{On a surface of constant
time, in our coordinates, the proper distance to the horizon from any finite
value of $r$ is infinite, but $r=0$ is reached for finite affine parameter on,
say, a timelike geodesic. Thus $r$ can be considered as a measure of distance
from the horizon.}.  The `end of the universe' limit
$\langle r\rangle\rightarrow\infty$ is also the free-field
limit so our results encompass previous free-field constructions
\cite{inthepast,A3} as well as previous partial constructions in which only the
field $r$ was considered \cite{Maldacena, KKR}.
\par
It seems likely that the methods used in
this paper can be extended to the full superconformal
theory. It may also be interesting to examine the
$(d,2)$ theories in greater detail.
Our discussion has been entirely at the level of classical field theory.
Whether the conformal invariance of brane actions exhibited here survives
quantization is an interesting open problem.
\medskip
\section*{Acknowledgments.}
\noindent
We had stimulating discussions with E. Bergshoeff, H.J.
Boonstra, M. Derix, M. Dine, M. Duff, S. Ferrara, E. Halyo,
J. Maldacena, A. Rajaraman and  E. Silverstein. The work of R.
K. and J. K. is supported by the NSF grant PHY-9219345.
The work of J. K. is also supported by the United States
Department of Defense, NDSEG Fellowship program. Work is
supported by the European Commission TMR programme
ERBFMRX-CT96-0045. We thank the theory division of CERN
for the hospitality and especially the organizers of the
workshop ``Non-Perturbative Aspects of Strings, Branes
and Fields", during which the first part of this work
was performed.
\medskip
\appendix
\section{The brane at the end of the universe}   \label{app:enduniverse}
One result of this paper is a one parameter class of static solutions of the
branewave equations in the $adS$ background provided by a supergravity
compactification. The parameter is a radial coordinate specifying the
location of the brane. Static solutions of this kind were also sought in
\cite{BDPS} but it was concluded there (for $p=2$) that such solutions exist only
when the brane is located at spatial infinity\footnote{Non-supersymmetric, and
non-static, solutions describing `finite radius' membranes were subsequently
found in \cite{A2} but we believe that these are different from the $p=2$
solutions described here. In particular, we expect our solutions to be
supersymmetric in the context of super $p$-brane actions. This point will be
addressed elsewhere.}. The purpose of this appendix is to explain this apparent
contradiction. Essentially, it is due to the fact that, because $adS$  space
admits more than one timelike Killing vector field, there is more than one
possible meaning to the word `static'. Each timelike Killing vector field
corresponds to a class of coordinate systems in which static solutions are
time-independent. Consequently, to understand why we find static solutions that
were not found in \cite{BDPS} requires an understanding of the differences
between the (horospherical) coordinates $(x^m,\phi)$ used here and the
coordinates used in \cite{BDPS}.
As before, we consider $adS_{p+2}$ as the hypersurface $T_1=0$ \eqn{defT1} in
${\hbox{\mybb E}}^{(p+1,2)}$ where the
latter has cartesian coordinates $(X^0, X^\pm, X^k)$ with $k=1,\dots,p$.
\par
Let for convenience
\begin{eqnarray}
X^0&=& T\nonumber\\
X^\pm &=& S\pm X^{p+1}   \nonumber\\
r^2 &=& \sum_{k=1}^p (X^k)^2 + (X^{p+1})^2 \ .
\end{eqnarray}
The constraint $T_1=0$ now becomes
\begin{equation}
T_1= -T^2 - S^2 + r^2 +(wR)^2=0\ .
\end{equation}
Whereas we solved the constraint in the main text by \eqn{horosphercoord},
we now consider (with $a=\frac1{wR}$)
\begin{eqnarray}
T&=& a^{-1} \sin (a\tau) (1 + a^2 r^2)^{\ft12}\,,\nonumber\\
S&=&a^{-1}\cos (a\tau) (1 + a^2 r^2)^{\ft12}\,,   \label{spher}
\end{eqnarray}
In these coordinates the metric is
\begin{eqnarray}
ds^2 &=&-dT^2-dS^2 +dr^2+ r^2 d\Omega_p^2\nonumber\\
&=& -(1 + a^2 r^2) d\tau^2 + (1+ a^2r^2)^{-1} dr^2 + r^2 d\Omega_p^2\ ,
\end{eqnarray}
where $d\Omega^2_p$ is the $SO(p+1)$-invariant metric on the $p$-sphere.
This is for $p=2$ the $adS_4$ metric of \cite{BDPS}.
\par
The Wess-Zumino term of
the M2 brane is determined by the 4-form \eqn{OmegaWZ}
\begin{equation}
\Omega_4 = - \frac {3a^2}{4!} \varepsilon_{\hat m_0 \dots\hat m_4} X^{\hat
m_0}dX^{\hat m_1} dX^{\hat m_2} dX^{\hat m_3}d X^{\hat m_4}\ .
\end{equation}
In the coordinates (\ref{spher}) with
\begin{eqnarray}
X_1 &=& r \cos \theta\,,\nonumber\\
X_2 &=& r \sin\theta\cos\phi\,,\nonumber\\
X_3 &=& r \sin\theta\sin\phi\,,
\end{eqnarray}
this becomes
\begin{equation}
\Omega_4 = - 3 a d\tau drd\theta d\phi (r^2 \sin\theta) = d(ar^3 \sin\theta
d\tau d\theta d\phi)\,,
\end{equation}
and we recover the $A_{012}$ of \cite{BDPS}.
\par
The Dirac action is
given by \eqn{Daction}.  In the gauge where we identify the world volume
coordinates with $\{\tau,\theta,\phi\}$, the field on the world-volume that
remains is $r(\tau,\theta,\phi)$. We can find the world-volume potential by
taking $r$ to be constant or all derivatives of $r$ vanishing.
The result for the world-volume potential is
\begin{equation}
V(r)=ar^3 - r^2 \sqrt{1+a^2r^2}\,.
\end{equation}
There is therefore a force proportional to
\begin{equation}
-V'(r) = 3ar^2 - 2r (1+a^2r^2)^{1\over2} + a^2 r^2 (1 + a^2
r^2)^{-{1\over2}}\, ,
\end{equation}
Except for $r=0$, this vanishes only for $r=\infty$ and we therefore recover
the result of \cite{BDPS} (see eq. (15) there). This yields the `membrane at the
end of the universe'.
The coordinates \eqn{spher} cover {\sl all} of $adS_{p+2}$ apart from $r=0$
which is (for $p>0$) the usual coordinate singularity at the origin. The
coordinate ranges are (except for $p=0$, in which case $r$ is unrestricted)
\be
0 \le \tau < 2\pi, \qquad 0 < r < \infty.
\ee
The time coordinate $\tau$ is identified with period $2\pi$. The global
structure is shown in the
Carter-Penrose diagram of Figure 1 in which each point is a $p$-sphere:
\begin{figure}[htb]
\begin{center}
\setlength{\unitlength}{.75mm}
\small
\begin{picture}(150,115)
\put(27,0){\epsfig{file=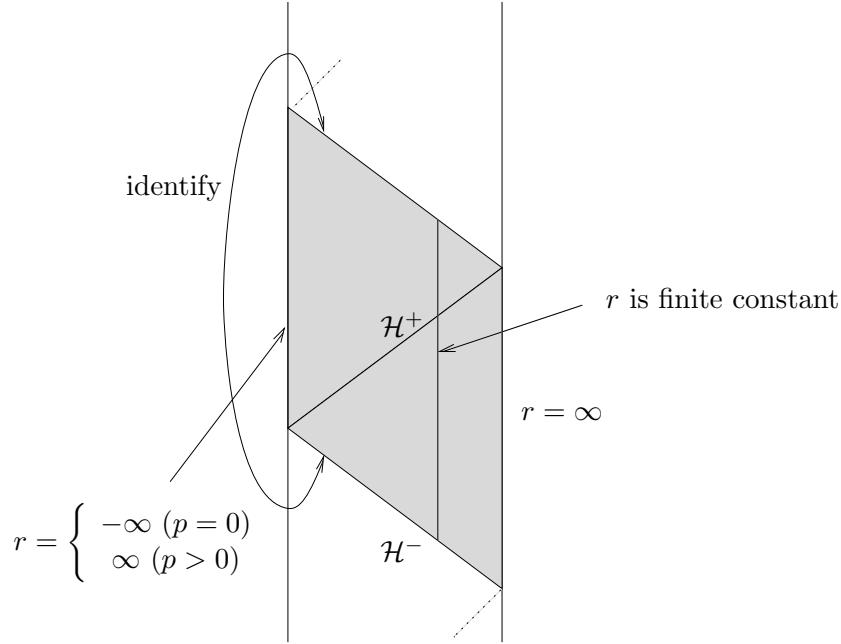,width=159pt}}
\put(65,55){${\cal H}^+$}
\put(65,15){${\cal H}^-$}
\put(90,40){$r=\infty$}
\put(20,80){identify}
\put(105,60){$r$ is finite constant}
\put(0,17){$r=\left\{\begin{array}{c}-\infty\ (p=0)\\ \infty\ (p>0)
\end{array}\right.$}
\end{picture}
\end{center}
\caption{Carter Penrose diagram with global structure in
parametisation $(\tau,r)$. Each point is a $p$-sphere.
The shaded region is $adS_{p+2}$.\label{fig:CP}}
\end{figure}
the shaded region in Fig.~\ref{fig:CP} is $adS_{p+2}$. Its universal cover is
obtained by dropping the identification under time translations, in which case
the diagram continues {\it ad infinitum} in both temporal directions. A
typical (constant but finite $r$) orbit of the timelike killing vector field
$\partial/\partial\tau$ is shown. Note that it necessarily cuts the null
hypersurfaces ${\cal H}^\pm$, the significance of which will become clear below. The
main result of \cite{BDPS} can be summarised by saying that the worldvolume field
configurations describing a static membrane (of topology $S^p$) at constant $r$
solves the relevant branewave equations only in the limit $r\rightarrow
\infty$. However, this does not mean that there cannot be solutions that are
static in terms of some other time coordinate.
\par
\begin{figure}[htb]
\begin{center}
\setlength{\unitlength}{.75mm}
\small
\begin{picture}(150,115)
\put(35,0){\epsfig{file=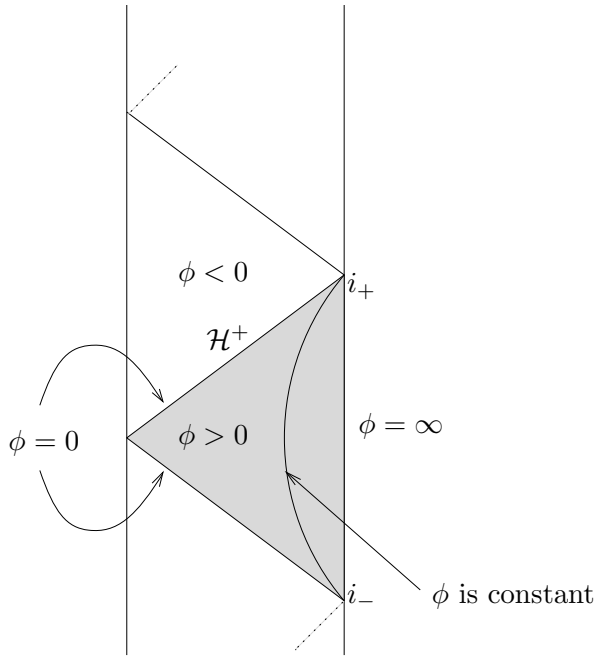,width=147pt}}
\put(60,67){$\phi<0$}
\put(65,55){${\cal H}^+$}
\put(92,40){$\phi=\infty$}
\put(60,38){$\phi>0$}
\put(30,37){$\phi=0$}
\put(105,10){$\phi$ is constant}
\put(90,10){$i_-$}
\put(90,65){$i_+$}
\end{picture}
\end{center}
\caption{Carter Penrose diagram with global structure in
horosperical coordinates.  The shaded region is covered by horospherical coordinates
for $\phi>0$.\label{fig:2}}
\end{figure}

In the horospherical coordinates used in the main text \eqn{horosphercoord},
$ds^2$ is singular at $\phi=0$, so that $\phi$ must be restricted to either the
`exterior' spacetime with $\phi >0$ or to the isometric `interior' spacetime
$\phi <0$. However, the singularity at $\phi=0$ is just a coordinate
singularity, as the previously introduced coordinates demonstrate. The
hypersurface $\phi=0$ is the union of a degenerate future Killing horizon (the
null hypersurface ${\cal H}^+$ marked on Fig.~\ref{fig:CP}) and a degenerate past Killing horizon
(${\cal H}^-$) of the timelike Killing vector field $\partial/\partial t$. The
`exterior' spacetime is shown in Fig.~\ref{fig:2} with a typical (constant but finite and
non-zero $\phi$) orbit of the timelike vector field $\partial/\partial t$.
As shown in the main text, a world-volume field configuration describing a static
membrane at constant $\phi$ solves the relevant branewave equations for {\sl any}
(non-zero) $\phi$. This is consistent with the results of \cite{BDPS} because a
constant $\phi$ hypersurface is static with respect to $\partial/\partial \tau$
only when the constant value of $\phi$ goes to infinity, and in this case the
constant $\phi$ hypersurface coincides with the hypersurface $r=\infty$, except
that the latter includes the points $i_\pm$ at infinity.
Note that the range of $t$ is $(-\infty,\infty)$ and that the proper time from
finite $t$ to $|t|=\infty$ on an orbit of $\partial/\partial t$ is infinite. A
similar state of affairs holds for the $\sigma$ coordinates, and hypersurfaces
of constant $\phi$ are globally isometric to $(p+1)$-dimensional Minkowski space.
This remains true for the limiting case in which $\phi=\infty$, so the
worldvolume of the `membrane at the end of the universe' is actually
$(p+1)$-dimensional Minkowski space. If the points at infinity implicit in the
$r=\infty$ description of this hypersurface are included then the worldvolume
is actually conformally compactified $(p+1)$-dimensional Minkowski space, which
is {\sl topologically} $S^1\times S^p$. We should point out (following
\cite{A3}, to which we refer for further references) that the mass terms
required for conformal field theories on $S^1\times S^p$ for finite radius
p-sphere are absent in the infinite radius limit, consistent with the absence of
such terms in the (interacting) conformal field theories found here.
\section{ Near-horizon supersymmetry of D1+D5} \label{app:susyD1D5}
Here we show that near the horizon the D1+D5 solution used in the main text
preserves 16 supersymmetries, which is double the number admitted by the full
D1+D5 solution of IIB supergravity. In the near-horizon geometry the dilaton is
a constant, which we may set to zero. Since the axion (pseudoscalar) and all
fermions also vanish, the IIB supersymmetry transformations reduce to\footnote{In
this section we use the mostly minus signature and the conventions given in
\cite{Berg}. }
\begin{eqnarray}
\delta \lambda &=& {3-p \over {4(p+2)!}} F_{{\mu}_1 ... {\mu}_{p+2}}
\gamma^{{\mu}_1 ... {\mu}_{p+2}} \imath \epsilon^{\ast }\label{l} \\
\delta \psi _{\mu}  &=& \partial_{\mu} \epsilon - {1\over 4} \omega
_{\mu} ^{\u a\u b}\gamma_{\u a} \gamma_{\u b} \epsilon +
{(-1)^{p} \over 8(p+2)!} F_{{\mu}_1 ... {\mu}_{p+2}}
\gamma ^{{\mu}_1 ... {\mu}_{p+2}} \gamma_{\mu}
\imath \epsilon^{\ast }  \label{psi} \ .
\end{eqnarray}
Filling in the near-horizon solution given in \eqn{onefive} it follows that
\be
\delta \lambda  = {\imath \gamma^{\u i}x_{i} \over R r}
\gamma^0 \gamma^5 (1-\gamma^{1234})
\epsilon ^{\ast} \ ,
\ee
where tangent-space indices are understood for the $\gamma$-matrices.
Preservation of supersymmetry requires $\delta\lambda=0$. We see that this is
satisfied by the 16 eigenspinors of $\gamma^{1234}$ with eigenvalue 1, i.e.
\be
(1 - \gamma^{1234})\epsilon^*_{kil} = 0\,, \label{constr}
\ee
To show that the D1+D5 solution admits 16 supersymmetries near
its horizon we must now show that spinors satisfying (\ref{constr})
also solve $\delta\psi_a=0$. In the limit $r\rightarrow 0$ and using
\eqn{constr}, it is straightforward to find
\begin{eqnarray}
\delta \psi_v &\approx& \partial_v \epsilon \,, \nonumber\\
\delta \psi_r &\approx& \partial_r \epsilon \,,\nonumber\\
\delta \psi_i &\approx& \partial_i \epsilon - \frac12 {x_i\over r^2} \epsilon
+ \frac12 \gamma^{\u j} \gamma_{\u i} {x_j \over r^2} \epsilon
- \frac i2 \gamma^{\u j} \gamma_{\u i} {x_j \over r^2} \gamma^0\gamma^5
\epsilon^*\,.
\end{eqnarray}
where
\be
v \in [0,5]
\qquad
r \in [1,2,3,4]
\qquad
i,j\in [6,7,8,9]\,.
\ee
Preservation of supersymmetry requires the variations of the gravitino
to vanish and we find an explicit solution for the killing spinors
\be
\epsilon_{kil} = \sqrt{\frac rR} (\epsilon_0 + i\gamma^0\gamma^5
\epsilon_0^*)\,,
\ee
where
\be
(1 - \gamma^{1234}) \epsilon_0 = 0
\ee
and $\epsilon_0$ constant.
Therefore near the horizon the D1+D5 configuration preserves 1/2 of
supersymmetry.
\section{Small velocity expansion} \label{app:smallv}
In this appendix we give the derivation of the expansion
of the world-volume actions for M2, D3 and M5, in the
limit of small velocities. We introduce flat indices
$\u\mu$, raised and lowered with $\eta_{\u\mu\u\nu}$. To
convert curved indices to flat ones we use a
\lq$\delta$-function vielbein', i.e.
\begin{equation}
T_{\mu\nu} = \delta_\mu^{\u\mu} \delta_\nu^{\u\nu}
T_{\u\mu\u\nu}\,.
\end{equation}
In the sequel these $\delta$-functions will be omitted.
The main difficulty is the expansion of
$$\det(\delta^\mu_{\ \nu} + X^\mu_{\ \nu})\,,$$ where $X^\mu_{\ \nu}$
is a matrix depending on the case considered.
The general form of the \lq Dirac' part, or generalizations thereof, of the
brane actions that we have considered is
\begin{equation}
I=I_D+I'\ ;\qquad
I_{D} = - \int d^{p+1} \sigma \left( \frac
rR\right)^{\ft{p+1}w}
\sqrt{\det (1 + X) }\,,
\end{equation}
with
\begin{eqnarray}
X^{\u\mu}_{\ \u\nu} &=& A^{\u\mu}_{\ \u\nu} + B^{\u\mu}_{\ \u\nu}\,,
\nonumber\\
A^{\u\mu}_{\ \u\nu} &=& \left( \frac Rr
\right)^{\ft{p+1}w}\partial^{\u\mu} X^{m'} \partial_{\u\nu}
X_{m'}\,,\label{matrixA}
\end{eqnarray}
where $B_{\bar \mu\bar \nu}$ is proportional to $F_{\mu\nu}$ or ${\cal H}^*_{\mu\nu}$,
hence antisymmetric. To go to the limit of small velocities we
note that $A$ is of order two in velocities and $B$ of
order one. $I'$ is only non zero for M5, see below.
Up to order four in velocities we have
\begin{eqnarray}
\tr X &\approx& \tr A\,, \nonumber\\
\tr X^2 &\approx& \tr A^2 + \tr B^2\,,\nonumber\\
\tr X^3 &\approx& 3 \tr (ABB)\,,\nonumber\\
\tr X^4 &\approx& \tr B^4\,.
\end{eqnarray}
Using the series-expansion for $\sqrt{\det(1+X)}$ we obtain,
\begin{eqnarray}
\sqrt{\det(1+X)}&=& 1 + \ft12 \tr A^2 - \ft14 \tr B^2\nonumber\\
& &-\ft14 [\tr A^2 - \ft12 (\tr A)^2] - \ft18 [\tr B^2 -
\ft14(\tr B^2)^2]\nonumber\\ & &+ \ft12 [\tr(ABB) - \ft14 \tr A
\tr B^2]\,,
\end{eqnarray}
up to fourth order in velocities.
So in all cases we find that
\begin{eqnarray}
I_{D} = - \int d^{p+1} x \left( \frac
rR\right)^{\ft{p+1}w}\hspace{-2mm}&& \Big(
1 + \ft12 \tr A - \ft14 \tr B^2\nonumber\\
& &-\ft14[\tr A^2 - \ft12 (\tr A)^2]\nonumber\\ &
&-\ft18[\tr B^4 -
\ft14 (\tr B^2)^2]\nonumber\\ & &+\ft12[\tr(ABB) - \ft14 \tr A \tr
B^2] + \dots \Big).\label{SBIWZ}
\end{eqnarray}
For M2 and D3 we have
\begin{eqnarray}
M2 & A^{\u\mu}_{\ \u\nu}=\left(\frac Rr\right)^6
\partial^{\u\mu} X^{m'} \partial_{\u \nu} X_{m'}& B^{\u\mu}_{\ \u\nu} =
0\,,\nonumber\\ D3 & A^{\u\mu}_{\ \u\nu}=\left(\frac
Rr\right)^4
\partial^{\u\mu} X^{m'} \partial_{\u \nu} X_{m'}& B^{\u\mu}_{\ \u\nu} =
\left(\frac Rr \right)^2 \eta^{\u\mu\u\rho}F_{\u\rho\u\nu}\,.
\end{eqnarray}
This yields the formulae of the main text.
The M5 brane has to be treated separately, due to the
many hidden metrics in the definitions of the fields
that appear in the action. But in the end everything
works out nicely. We take $w=2$. The inverse metric
\begin{eqnarray}
g^{\mu\nu} &=& \left( \frac Rr \right)
(\delta^{\u\mu}_{\
\u\rho} +
\sum_{n=1}^\infty (-)^n (A^n)^{\u\mu}_{\ \u\rho}) \eta^{\u\rho\u\nu}
\nonumber\\ &\approx& \left( \frac Rr \right) (\eta^{\u\mu\u\nu}
- A^{\u\mu\u\nu} + \dots )
\end{eqnarray}
up to second order in velocities (note that since this
metric is used to contract indices of fields we only
need it to this order)\,. $A^{\u\mu}_{\ \u\nu}$ is given
in (\ref{matrixA}). Taking the definitions in (5.3) of
\cite{m5tens} and expanding the metrics, we have
\begin{eqnarray}
\sqrt g &\approx& 1 + \ft12 A^{\u \mu}_{\ \u\mu}\,, \nonumber\\
u^2 &=& u_\mu g^{\mu\nu} u_\nu \approx \left( \frac Rr
\right) (u_{\u
\mu} \eta^{\u\mu\u\nu} u_{\u\nu} - u_{\u\mu} A^{\u\mu\u\nu}
u_{\u\nu})\,,\nonumber\\ u_{\u\mu}&=& \partial_{\u\mu}
a\,,\nonumber\\ v_\rho &\equiv& \frac
{u_\rho}{\sqrt{u^2}}
\approx
\left( \frac rR \right)^{1/2} v_{\u \rho} ( 1 + \ft12 v_{\u\mu}
A^{\u\mu\u\nu} v_{\u\nu})\,,\nonumber\\
\H^{*\mu\nu} &=& \frac 1{6 \sqrt g}
\varepsilon^{\mu\nu\rho\sigma\tau\phi}
v_\rho \H_{\sigma\tau\phi}
\approx \left( \frac Rr \right)^{5/2} \H^{*\u\mu\u\nu} ( 1 + \ft12
v_{\u\rho} A^{\u\rho\u\sigma} v_{\u\sigma} - \ft12
A^{\u\rho}_{\
\u\rho})
\,,\nonumber\\
\H^*_{\mu\nu}
&\approx& \left( \frac Rr \right)^{1/2}
(\H^{*\u\mu\u\nu} ( 1 +
\ft12 v_{\u\rho} A^{\u\rho\u\sigma} v_{\u\sigma} - \ft12
A^{\u\rho}_{\ \u\rho}) + A_{\u\mu\u\rho} \H^{*\u\rho}_{\
\
\u\nu} +
\H^*_{\u\mu\u\rho} A^{\u \rho}_{\ \u\nu})\,, \nonumber\\
\H_{\mu\nu} &\equiv& g^{\rho\sigma} v_{\rho} \H_{\sigma\mu\nu}
\approx \left( \frac Rr \right)^{1/2} (\H_{\u\mu\u\nu} - v_{\u\rho}
A^{\u\rho\u\sigma} \H_{\u\sigma\u\mu\u\nu} + \ft12
v_{\u\rho} A^{\u\rho\u\sigma} v_{\u \sigma}
\H_{\u\mu\u\nu})\,.\nonumber\\
\end{eqnarray}
On the
right hand side of the $\approx$-sign every index is contracted with
$\eta_{\u\mu\u\nu}$.
For $I'$, which is the last term of \eqn{M5action}, this gives
\begin{eqnarray}
I' &=& - \frac14 \int d^6 \sigma \ \sqrt{-\det g} \H^{*\mu\nu}
\H_{\mu\nu}\nonumber\\
&\approx& - \frac14 \int d^6 \sigma \H^{*\u\mu\u\nu}
\H_{\u\mu\u\nu}\nonumber\\ &&
\hspace{2cm} - \frac14 \H^{*\u\mu\u\nu}(v_{\u\rho} A^{\u\rho\u\sigma}
v_{\u\sigma}
\H_{\u\mu\u\nu} - v_{\u\rho} A^{\u\rho\u\sigma}
\H_{\u\sigma\u\mu\u\nu})\,.
\label{SH}
\end{eqnarray}
Now we can combine everything and use formula
(\ref{SBIWZ}) with
\begin{eqnarray}
A^{\u\mu}_{\ \u\nu} &=& \left( \frac Rr \right)^3
\partial^{\u\mu} X^{m'}
\partial_{\u\nu} X_{m'}\,,\nonumber\\
B^{\u\mu}_{\ \u \nu}&=& i \left( \frac Rr \right)^{3/2}
\Big[\H^{*\u\mu}_{\
\ \u\nu} ( 1
+ \ft12 v_{\u\rho} A^{\u\rho\u\sigma} v_{\u\sigma} -
\ft12 A^{\u\rho}_{\ \u\rho}) + A^{\u\mu\u\rho}
\H^*_{\u\rho\u\nu} +
\H^{*\u\mu\u\rho} A_{\u\rho\u\nu}\Big]\,,\nonumber\\
\end{eqnarray}
together with (\ref{SH}) to obtain the formula of the
main text.


\medskip
\begin{thebibliography}{99}
\bibitem{Fronsdal}
C. Fronsdal, {\it Singletons}, in proceedings of the 1989 {\sl Latin American
School on Strings and Fundamentals}, preprint UCLA/89/TEP/66.
\bibitem{BDPS} E. Bergshoeff, M.J. Duff, C.N. Pope and E. Sezgin, {\it
Supersymmetric supermembrane vacua and singletons},
Phys. Lett. {\bf 199B} (1987) 69; {\it Compactifications of the
eleven-dimensional supermembrane}, Phys. Lett. {\bf 224B} (1989) 71.
\bibitem{GT} G. W. Gibbons and P. K. Townsend, {\it Vacuum
interpolation in supergravity via super p-branes}, Phys.
Rev. Lett. {\bf 71} (1993) 3754 ; hep-th/9307049.
\bibitem{inthepast}
H. Nicolai and E. Sezgin, {\it Singleton representations of $OSp(N|4)$}, Phys.
Lett. {\bf 143B} (1984) 389; \\
M. G{\" u}naydin, B.E.W. Nilsson, G. Sierra and P.K. Townsend, {\sl Singletons
and superstrings}, Phys. Lett. {\bf 176B} (1986) 45;\\
E. Bergshoeff, A. Salam, E. Sezgin and Y. Tanii, {\it N=8
supersingleton quantum field theory}, Nucl. Phys. {\bf B305} (1988) 497; \\
H. Nicolai, E. Sezgin and Y. Tanii, {\it Conformally invariant supersymmetric
field theories on $S^p\times S^1$ and super p-branes}, Nucl. Phys. {\bf B305}
(1988) 483.
\bibitem{A3}
M.P. Blencowe and M.J. Duff, {\sl Supersingletons}, Phys. Lett. {\bf 203B}
(1988) 229.
\bibitem{m5tens} P. Claus, R. Kallosh and A. Van Proeyen, {\it M
5-brane and superconformal (0,2) tensor multiplet in 6
dimensions}, Nucl. Phys. {\bf B}, to be published, hep-th/9711161.
\bibitem{Maldacena}
J. Maldacena, {\it  The large N limit of superconformal field theories and
supergravity}, hep-th/9711200.
\bibitem{KKR}
R. Kallosh, J. Kumar and  A. Rajaraman, {\it Special conformal symmetry of
worldvolume actions}, hep-th/9712073.
\bibitem{FF} S. Ferrara and C. Fronsdal, {\it Conformal Maxwell theory
as a singleton field theory on $adS_5$, $IIB$ three branes and duality},
hep-th/9712239.
\bibitem{DGT}
M.J. Duff, G.W. Gibbons and P.K. Townsend, {\it Macroscopic superstrings as
interpolating solitons}, Phys. Lett. {\bf 332 B} (1994) 321; \\
G.W. Gibbons, G.T. Horowitz and P.K. Townsend, {\it Higher-dimensional
resolution of dilatonic black hole singularities}, Class. Quantum Grav.
{\bf 12} (1995) 297.
\bibitem{vafa}
C. Vafa, {\it Evidence for F-theory}, Nucl. Phys. {\bf B469} (1996) 403,
hep-th/9602022;\\
 C.M. Hull, {\it String dynamics at strong coupling}, Nucl. Phys. {\bf B468}
(1996) 113, hep-th/9512181;\\
S. Hewson, {\it An approach to F-theory}, hep-th/9712017.
\bibitem{bars}
I. Bars, {\it S-theory},  Phys. Rev. {\bf D55} (1997) 2373, hep-th/9607112.
\bibitem{sezgin}
I. Rudychev, E. Sezgin and P. Sundell, {\it Supersymmetry in dimensions beyond
eleven}, hep-th/9711127.
\bibitem{duff} M. Blencowe and M.J. Duff, {\it Supermembranes and
signature of spacetime}, Nucl. Phys. {\bf B310} (1988) 387;\\
D. Kutasov and E. Martinec, {\it M-branes and N=2 strings}, Class. Quantum Grav.
{\bf 14} (1997) 2483;\\
S. Hewson and M. Perry, {\it The Twelve dimensional super (2+2)-brane},
Nucl.Phys. {\bf B492} (1997), 249, hep-th/9612008.
\bibitem{Hyun} S. Hyun, {\it U-duality between Three and Higher Dimensional Black Holes},
hep-th/9704005.
\bibitem{KK} R.~Kallosh and J.~Kumar, {\it Supersymmetry enhancement
 of $D$-$p$-branes and $M$-branes}, Phys. Rev. {\bf D56} (1997) 4934,
hep-th/9704189.
\bibitem{Skenderis}
H.J.~Boonstra, B.~Peeters and K.~Skenderis,
{\it Duality and asymptotic geometries}, Phys. Lett.
{\bf 411B} (1997) 59, hep-th/9706192; {\it Branes and
anti-de Sitter space-times},
to appear in the proceedings of the conference ``Quantum Aspects
of Gauge theories, Supersymmetry and Unification'', Neuchatel,
September 1997, hep-th/9801076; K. Sfetsos
and K. Skenderis, {\it Microscopic derivation of the
Bekenstein--Hawking entropy formula for non-extremal
black holes}, Nucl. Phys. {\bf B} to be published, hep-th/9711138.
\bibitem{cowtown}
P.M. Cowdall and P.K. Townsend, {\it Gauged supergravity vacua from
intersecting branes}, hep-th/9801165.
\bibitem{nahm}
W. Nahm, {\it Supersymmetries and their representations}, Nucl. Phys. {\bf B135}
(1978) 149.
\bibitem{GNST}
M. G{\" u}naydin, G. Sierra and P.K. Townsend, {\sl The unitary supermultiplets
of d=3 anti-de Sitter and d=2 conformal superalgebras}, Nucl. Phys. {\bf B274}
(1986) 429;\\
M. G{\" u}naydin, B.E.W. Nilsson, G. Sierra and P.K. Townsend, {\sl Singletons
and superstrings}, Phys. Lett. {\bf 176B} (1986) 45.
\bibitem{CLLPS} E. Cremmer, I.V. Lavrinenko, H. Lu, C.N. Pope, K.S. Stelle and T.A. Tran,
{\it Euclidean signature supergravities, dualities and instantons},  hep-th/9803259.
\bibitem{us}  P. Claus, M. Derix, R. Kallosh, J. Kumar, P.K. Townsend
and A. Van Proeyen,
{\it Black holes and superconformal mechanics}, hep-th/9804177.
\bibitem{5b} I. Bandos, K. Lechner, A. Nurmagambetov, P. Pasti, D.
Sorokin and M. Tonin, {\it Covariant action for the
super-five-brane of M-theory}, Phys. Rev. Lett. {\bf 78}
(1997) 4332.
\bibitem{CM} C.G.~Callan and J.M.~Maldacena, {\it D-brane approach to black
hole quantum mechanics},
 Nucl.~Phys. {\bf B472} (1996)
591, hep-th/9602043.
\bibitem{tsey} A. Tseytlin, {\it Extreme dyonic black holes in string theory},
Mod. Phys. Lett. {\bf A11} (1996) 689, hep-th/9601177.
\bibitem{BGS} Fr. Brandt, J. Gomis and J. Sim\'on, {\it The rigid symmetries of
bosonic D-strings}, hep-th/9803196.
\bibitem{A2}
E. Bergshoeff, M.J. Duff, C.N. Pope and E. Sezgin, {\sl Compactifications of
the eleven-dimensional supermembrane}, Phys. Lett. {\bf 224B} (1989) 71.
\bibitem{Berg} E. Bergshoeff, {\it P-Branes, D-Branes and
M-Branes}, in `Gauge Theory
      Applied Supersymmetry and Quantum Gravity II', eds.
      A. Sevrin, K.S. Stelle, K. Thielemans and A. Van Proeyen,  Imperial
College Press, p. 210, hep-th/9611099.
\end{thebibliography}
\end{document}